\documentstyle[12pt]{article}
\begin{document}
\setlength{\unitlength}{0.75mm}\thicklines
\begin{flushright}
Preprint IHEP 94-77\\
September 1994\\
Submitted to Phys.Rev.D.
\end{flushright}

\begin{center}
{\bf \large Semileptonic $\boldmath{B\to D^{(*)}l\nu}$ decays, the slope of
Isgur-Wise function and \\
$\boldmath{|V_{bc}|}$ value in potential quark model}
\\
\vspace*{1cm}
V.V.Kiselev\\
{\it Institute for High Energy Physics,\\
Protvino, Moscow Region, 142284, Russia,\\
E-mail: kiselev@mx.ihep.su\\
Fax: +7-095-230-23-37}
\end{center}

\begin{abstract}
The modification of the naive potential model of the heavy meson is considered
on the basis of the single-time prescription for the quark-meson form factor
and the covariant choice of the quatk relative momentum. The application of
the model to the description of the semileptonic decays of the beauty
meson allows one to determine the Isgur-Wise function slope $\rho^2\simeq 1.25$
and to extract the value $|V_{bc}|= 0.038\pm 0.003$ from the experimental data.
\end{abstract}

\section{Introduction}

The extraction of a value for the matrix element of the weak charged current
mixing $|V_{bc}|$ from the data on the semileptonic $B\to D^{(*)}l\nu$ decays
demands the modelling of the nonperturbative dynamics of QCD in the framework
of the sum rules \cite{1,2} or the potential quark models \cite{3,4}.
The detailed experimental researches of the differential characteristics
for the $B\to D^{(*)}l\nu$ decays \cite{5,6} have shown, that the
measured slope of the Isgur-Wise function $\rho^2= -\xi'_{IW}(y=1)$ \cite{7}
does not agree with the estimate in the naive potential model \cite{3}
(see also ref.\cite{8})
\begin{equation}
\rho^2_{naive} = \frac{m_{sp}^2}{\omega^2} \simeq 0.6\;, \label{1}
\end{equation}
where $m_{sp}$ is the mass of the spectator-quark, $\omega$ is the parameter,
determining the width of the wave function, having the oscillator type,
$\phi(\vec{q}) \sim \exp(-\vec{q}^2/\omega^2)$, when in the fitting of the
experimental data the different $\xi_{IW}(y)$ parametrizations give \cite{2,9}
\begin{equation}
\rho^2 = 1.2\pm 0.3\;.
\end{equation}
Naive estimate (\ref{1}) does not also agree with the values, obtained in the
QCD sum rules \cite{2} and in the lattice computations \cite{10}
\begin{eqnarray}
\rho^2_{SR}& = & 1.0\div 1.3\;,\\
\rho^2_{Lat}& = & 1.24\pm 0.26\pm 0.33\;.
\end{eqnarray}

In the original ISGW paper \cite{3}, coming from the experimental data on
the elastic form factor of the pion, the authors have introduced
the correcting factor $k \simeq 0.7$, corresponding to account of relativistic
corrections for the light spectator quark, so that
\begin{equation}
\rho^2_{ISGW} = \frac{m_{sp}^2}{\omega^2}\; \frac{1}{k^2} \simeq 1.2\;.
\end{equation}
In paper \cite{11} one has noted, that the $k$-factor appearance is mainly
caused by the accurate covariant choice of the relative momentum of quarks
inside the recoil meson at its nonzero momentum in the rest frame of the
decaying meson.

In the mentioned potential models, the mesonic state is built up as the
superposition of free constituent quarks, so that this wave packedge belongs
to the single-particle representation of the Poincare group in the limit of
the infinitely narrow wave function ($\omega\to 0$) only.

In ref.\cite{12} the covariant form of the potential quark model has been
offered on the basis of the single-time prescription for the quark-meson
form factor.

In the present paper we make the systematic consideration of form factors
for the semileptonic decays of heavy mesons in the covariant model with the
account of the relative quark momentum modification, offered in ref.\cite{11}.
So, we find, that in the potential model with the wave function of the
oscillator form, the optimal value of the Isgur-Wise function slope is
equal to
\begin{equation}
\rho^2 = 1.25\;,
\end{equation}
so that the fit of the experimental $d\Gamma(B\to D^*l\nu)/dq^2_{l\nu}$
spectrum leads to the result
\begin{equation}
|V_{bc}| = 0.038\pm 0.003\;.
\end{equation}

\section{Quark-meson Form Factor}

The nonrelativistic wave function for a bound $S$-state of quarks with
the masses $m$ and $m'$ in the meson rest frame has the form
\begin{equation}
\Psi(t,\; \vec{x},\; \vec{x}') = e^{-i(m+m'+\epsilon)t}\;
\phi(\vec{x}-\vec{x}')\; C^J_{SS'}\;, \label{9}
\end{equation}
where $\epsilon$ is the binding energy of the quarks, $C$ is the spin factor.
Function (\ref{9}) corresponds to the
quark-meson form factor with the single-time prescription
\begin{equation}
\chi(P;q) = 2\pi\;\delta(Pq)\; \bar \phi (\bar q^2)\;, \label{10}
\end{equation}
so that the momenta of quarks, entering the meson, are equal to
\begin{eqnarray}
k = \frac{m}{m+m'}\;P + q\;, \nonumber \\
k' = \frac{m'}{m+m'}\;P - q\;, \label{11}
\end{eqnarray}
and the relative momentum $\bar q$ is defined by the expression \cite{11}
\begin{equation}
\bar q = q - P\frac{(P,q)}{M^2}\;,
\end{equation}
where $P$ is the meson momentum, $P^2=M^2$, $N$ in eq.(\ref{10}) is the
normalization factor. So, the quark meson vertex has the form
\begin{equation}
L_{q\bar q' M} = \bar u(-k) \Gamma_M u'(k')\;D^{-1}(k)\;D^{-1}(k')\;
\chi (P;q)\;, \label{13}
\end{equation}
and
\begin{equation}
\Gamma_P = \gamma_5\; \frac{1+v_\mu\gamma^\mu}{2}
\end{equation}
for the pseudoscalar $0^-$-state, and
\begin{equation}
\Gamma_V = \epsilon_{\mu} \gamma^{\mu}\; \frac{1+v_\mu\gamma^\mu}{2}
\end{equation}
for the vector $1^-$-state with the polarization vector $\epsilon_{\mu}$.
The $D(k)$ value defines the quark propagator
\begin{equation}
S(k) = (k_{\mu} \gamma^{\mu} + m)\;D(k)\;, \label{16}
\end{equation}
so, in the limit of a low energy for the quark binding $(\epsilon \ll m$),
we have
\begin{equation}
Im\;D(k) = \frac{\pi M}{m}\;\delta (Pq) \;. \label{17}
\end{equation}
One can easily show, that in the single-time prescription, form factor
(\ref{10}) satisfies the Bethe-Salpeter equation in the leading order
over the low binding energy
\begin{equation}
\Gamma_M\; \chi(P;q) = D(k) D(k') \int \frac{d^4q'}{(2\pi)^4}\;
V(\bar q')\; \Gamma_M\; \chi(P;q')\; 4\mu M\;, \label{18}
\end{equation}
after the integration over $q_0$ in the frame of $P=(M,\vec{0})$, if
$D(k)$ is approximated by the expression for the free quark and $\bar \phi
(\bar q)$ satisfies the Schr\" odinger equation with the potential
$V(\bar q)$ and the reduced mass $\mu$.

One has to note, that the constituent masses of quarks are defined with
the accuracy up to a low value of the order of the binding energy.

Further, the leptonic constants are determined by the expression\footnote{
We use the normalization with $f_\pi \simeq 132$ MeV.}
\begin{equation}
f=f_{P,V} = 2 \sqrt{\frac{3}{M}}\;\phi (0)\;, \label{19}
\end{equation}
where $\phi(0)$ is the normalized wave function at the origin.
In the consideration of the oscillator function
\begin{equation}
\phi (\vec{r}) = \biggl(\frac{\omega^2}{2 \pi}\biggr)^{3/4}\;
\exp(-r^2\omega^2/4)\;, \label{20}
\end{equation}
we will have
\begin{eqnarray}
\bar \phi(\bar q) & = & \exp(\bar q^2/\omega^2)\;\\
N & = & \frac{M}{m m'}\; \frac{\sqrt{6}}{f}\;.
\end{eqnarray}
Further, in the ISGW  model \cite{3} the $\omega$ parameter was determined
from the variation principle, when one found the best description of the heavy
meson spectroscopy in the Cornel model potential \cite{13} in the class of the
probe functions of the oscillator form, since, generally, with the
increase of the distance between the quarks, the nonrelativistic
wave functions in the different potential models are rapidly dropping
under the law, close to the exponential one.

In the present paper we use another procedure for the choice of the $\omega$
parameter.

Indeed, approximating the heavy $(Q\bar q)$ meson wave function in the
oscillator form (see eq.(\ref{20})) with the accuracy up to $O(1/m_Q)$-terms,
corresponding to the spin-spin splitting, for example, for meson mass we
get the expression
\begin{equation}
M= m_Q+m_q+ \frac{3}{4}\; \frac{\omega^2}{m_q} +\delta_V\;,
\end{equation}
where $\delta_V$ is the contribution of a perturbation, corresponding to
the deviation of potential from the oscillator one. Then, let us chose the
model parameters, so that it would describe the meson mass by the best
way, i.e. we suppose
\begin{equation}
\delta_V =0\;.\label{24}
\end{equation}
Further, the effective mass of the constituent light quark must, by its sense,
be determined by the dimensional parameter of the meson wave function.
Thus, the only parameter of the model with the oscillator wave function must
be the $\omega$ quantity, and the quantity
\begin{equation}
\bar \Lambda = M- m_Q
\end{equation}
must not depend on $m_q$, i.e.\footnote{
We neglect the contribution of the low current mass of the light quark.}
\begin{equation}
\frac{\partial \bar \Lambda}{\partial m_q} = 0\;. \label{26}
\end{equation}
{}From eqs.(\ref{24}) and (\ref{26}) it follows, that
\begin{eqnarray}
m_q(\omega) & = & \frac{\sqrt{3}}{2} \omega\;,\\
\bar \Lambda & = & \sqrt{3} \omega\;.
\end{eqnarray}
Thus, the $\bar \Lambda$ quantity is the only parameter of the model in the
leading approximation over the inverse heavy quark mass, so that
$m_Q=M-\bar \Lambda$.

The $\bar \Lambda$  estimates have been obtained in the QCD sum rules for
both the heavy mesons \cite{2,14}
\begin{equation}
\bar \Lambda = 0.57\pm 0.07\;\;GeV\;,
\end{equation}
and the heavy quarkonium \cite{15}
\begin{equation}
\bar \Lambda = 0.59\pm 0.02\;\;GeV\;. \label{30}
\end{equation}
In the furthercoming calculations we accept the mean value in eq.(\ref{30}).
Then, for the leptonic constants of $B$- and $D$-mesons, we obtain
the estimates
\begin{eqnarray}
f_B & \simeq & 80\;\; MeV\;,\\
f_D & \simeq & 130\;\; MeV\;,
\end{eqnarray}
which agree with the estimates in the QCD sum rules \cite{1}, where $f_B=
90\div 200$ MeV, $f_D = 120\div 250$ MeV. For the mass of the light
constituent quark, we have
\begin{equation}
m_q \simeq 0.3\;\; GeV\;,
\end{equation}
that is also in a good agrement with the notions about this parameters
in the quark model of hadrons.

\section{Semileptonic Form Factors}

Let us consider the amplitude of the $B\to D^{(*)}l\nu$ transition (see
Fig.\ref{f1})
\begin{equation}
A = \frac{G_F}{\sqrt{2}}\;V_{bc}\;l_{\mu}\;H^{\mu}\;, \label{34}
\end{equation}
where $l_\mu$ is the weak current of leptons, $H^\mu$ is the hadronic matrix
element of the $J_\mu$ current
\begin{equation}
J_{\mu} = V_\mu - A_\mu = \bar b \gamma_{\mu} (1 - \gamma_5) c\;.
\end{equation}
Introduce the standard definitions of the scalar form factors
\begin{eqnarray}
<B(p)| A_{\mu} |D(k)> & = & F_+(t) (p+k)_{\mu} + F_-(t) (p-k)_{\mu}\;,
       \label{36} \\
<B(p)| J_{\mu} |D^*(k,\lambda)>
& = & -(M+M_{D^*}) A_1(t)\;\epsilon_{\mu}^{(\lambda)} \nonumber \\
& ~ & +  \frac{A_2(t)}{M+M_{D^*}}\;
     (\epsilon^{(\lambda)} p)\;(p+k)_{\mu} \nonumber \\
& ~ &  + \frac{A_3(t)}{M+M_{D^*}}\;
     (\epsilon^{(\lambda)} p)\;(p-k)_{\mu}  \nonumber \\
& ~ &  + i \frac{2 V(t)}{M+M_{D^*}}\;
     \epsilon_{\mu \nu \alpha \beta}\;
     \epsilon_{(\lambda)}^{\nu} p^\alpha k^\beta\;. \label{37}
\end{eqnarray}
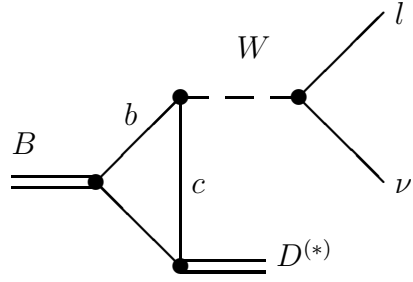
\begin{figure}[t]
\begin{center}
\begin{picture}(90,60)
\put(25,20){\circle*{3}}
\put(40,5){\circle*{3}}
\put(40,35){\circle*{3}}
\put(25,20){\line(1,1){15}}
\put(40,35){\line(0,-1){30}}
\put(40,5){\line(-1,1){15}}
\put(10,21){\line(1,0){15}}
\put(10,19){\line(1,0){15}}
\put(40,6){\line(1,0){15}}
\put(40,4){\line(1,0){15}}
\put(40,35){\line(1,0){5}}
\put(48,35){\line(1,0){5}}
\put(56,35){\line(1,0){5}}
%\put(64,35){\line(1,0){5}}
%\put(40,35){\line(1,1){3}}
%\put(43,38){\line(1,-1){6}}
%\put(49,32){\line(1,1){6}}
%\put(55,38){\line(1,-1){6}}
%\put(61,32){\line(1,1){3}}
\put(61,35){\line(1,1){15}}
\put(61,35){\line(1,-1){15}}
\put(61,35){\circle*{3}}

\put(10,25){$B$}
\put(78,48){$l$}
\put(78,18){$\nu$}
\put(50,42){$W$}
\put(30,30){$b$}
\put(42,18){$c$}
\put(57,5){$D^{(*)}$}
\end{picture}
\end{center}
\caption{Semileptonic $B$ decay due to $b$ transition into $c$.}
\label{f1}
\end{figure}

Then in the model under the consideration, we get the following explicit
expressions at $t=(p-k)^2$
\begin{eqnarray}
F_+(t) & = & \frac{1}{2}\; (m_b+m_c) \sqrt{\frac{M_D}{M}}
        \frac{1}{m_c} \xi_D (t)\;, \nonumber\\
F_-(t) & = & -\frac{1}{2}\; (m_b-m_c+2 m_{sp}) \sqrt{\frac{M_D}{M}}
        \frac{1}{m_c} \xi _D (t)\;, \nonumber \\
V(t) & = & \frac{1}{2}\; (M+M_{D^*}) \sqrt{\frac{M_{D^*}}{M}}
        \frac{1}{m_c} \xi _{D^*} (t)\;, \label{38} \\
A_1(t) & = & \frac{1}{2}\; \frac{M^2+M_{D^*}^2-t+2 M (m_c-m_{sp})}{M+M_{D^*}}
        \sqrt{\frac{M_{D^*}}{M}} \frac{1}{m_c} \xi _{D^*} (t)\;,\nonumber \\
A_2(t) & = & \frac{1}{2}\; (M+M_{D^*}) (1-2 m_{sp}/M) \sqrt{\frac{M_{D^*}}{M}}
        \frac{1}{m_c} \xi _{D^*} (t)\;, \nonumber \\
A_3(t) & = & -\frac{1}{2}\; (M+M_{D^*}) (1+2 m_{sp}/M) \sqrt{\frac{M_{D^*}}{M}}
        \frac{1}{m_c} \xi _{D^*} (t)\;, \nonumber
\end{eqnarray}
where at $X=D,\; D^*$
\begin{eqnarray}
\xi _X(t) & = & \frac{2 \omega_B \omega_X}{\omega_B^2+\omega_X^2}
\sqrt{\frac{2 \omega_B \omega_X}{\omega_B^2 y^2+\omega_X^2}}\; \nonumber\\
&& \exp\biggl(-\frac{m_{sp}^2}{\omega_B^2 y^2+\omega_X^2}\; (y^2-1)\biggr)\;,
\end{eqnarray}
with $y=(p,k)/M M_X$, so that
$$
y-1 = \frac{t_m-t}{2 M M_X}\;,
$$
where $t_m=(M-M_X)^2$ is the maximum square of the lepton pair mass.

The expressions for the differential widths $d\Gamma/dk_X$, where $k_X$ is
the momentum of the recoil meson in the $B$-meson rest frame,
can be found in ref.\cite{12}, for instance.

Note, that the account of the modification of the relative quark momentum
in the meson has been also made in ref.\cite{17}, where one has considered the
$B_c$-meson decays. However, first, one has used the kinematics
of free quarks (this has made the consideration to be very bulky), and, second,
one has substituted implicit numerical expressions for the heavy quarkonium
wave functions, so this does not allow one to make an explicit comparison
with the oscillator function models, used wide.

\section{Infinitely Heavy Quark Limit}

As is shown in ref.\cite{7}, in the limit $m_{b,c}\gg \Lambda_{QCD}$, the
matrix elements of the weak currents take the universal form
\begin{eqnarray}
<B(p)| J_{\mu} |D^*(k,\lambda)> & = &\sqrt{M M_{D^*}}\; \xi_{IW}(v\cdot v')\;
      \times \nonumber\\ & ~ & \times
      (\epsilon_\mu (1+v\cdot v') - v'_\mu (e\cdot v)- i
      \epsilon_{\mu\nu\alpha\beta} v^\nu v'^\alpha \epsilon^\beta)\;,
      \label{41} \\
<B(p)| J_{\mu} |D(k)> & = &\sqrt{M M_{D}}\;
      \xi_{IW}(v\cdot v')\; (v + v')_\mu\;,
\end{eqnarray}
where $v=P/M$, $v'=k/M_X$, $y=v\cdot v'$. In this limit, the wave functions
of the heavy mesons do not depend on the heavy quark flavour
\begin{equation}
\omega = \omega_B = \omega_{D^{(*)}}\;,
\end{equation}
i.e. the scaling law for the leptonic constants takes place
\begin{equation}
f^2\cdot M = const.\;, \label{44}
\end{equation}
and the following condition is satisfied
\begin{equation}
\xi_{IW}(y=1) = 1\;. \label{45}
\end{equation}
In the model, being considered, the infinitely heavy quark
limit leads to the validity of relations (\ref{41})-(\ref{45}) and to the
expression
\begin{equation}
\xi (y) = \sqrt{\frac{2}{1+y^2}}\;
\exp\biggl(-\frac{m_{sp}^2}{\omega^2 (1+y^2)}\;(y^2-1)\biggr)\;, \label{46}
\end{equation}
that, in the model parameter optimization considered above, results in
\begin{equation}
\xi_{opt} (y) = \sqrt{\frac{2}{1+y^2}}\;
\exp\biggl(-\frac{3}{4}\; \frac{y^2-1}{y^2+1}\biggr)\;. \label{47}
\end{equation}
At the low $(y-1)$ values, eq.(\ref{46}) gives
\begin{equation}
\xi (y) = 1 - \biggl(\frac{1}{2}+\frac{m_{sp}^2}{\omega^2}\biggr) (y-1)\;,
\label{48}
\end{equation}
or in the optimal regime, one has
\begin{equation}
\xi (y) = 1 - 1.25 (y-1)\;, \;\;\; \rho^2 = -\xi'(y=1) =1.25\;.
\label{49}
\end{equation}

\section{Meson Dominance for the Form Factors}

In accordance with the meson dominance, the form factors of
the semileptonic meson transitions must have the form
\begin{equation}
F(t) = \frac{F(0)}{1-t/M_P^2}\;, \label{50}
\end{equation}
where $M_P$ is the mass of the meson, corresponding to the transition
current. For the semileptonic $B\to D^{(*)}l\nu$ decays, one has
$M_P= M_{B_C}= 6.3\div 6.8$ GeV \cite{18}. In the limit
$m_{b,c}\gg \Lambda_{QCD}$, one will have $M_P=(m_b+m_c)$, and
dependence (\ref{50}) results in
\begin{equation}
\xi_{MD}(y) = \frac{2}{y+1}\;, \label{51}
\end{equation}
so that
\begin{equation}
\rho^2_{MD} = \frac{1}{2}\;. \label{52}
\end{equation}
{}From eqs.(\ref{48}) and (\ref{52}) it follows, that in the quark model, the
meson dominance takes place under the condition
$$
m_q \ll \omega\;,
$$
so that at $m_q\to 0$
$$
\frac{3}{4}\; \frac{\omega^2}{m_q} \simeq \bar \Lambda \ll m_Q\;,
$$
i.e. only when the wave packedge is infinitely narrow in comparison
with the heavy quark mass and, instanteniously, it is infinitely broad in
comparison with the constituent mass of the light quark. However,
these conditions are explicitly not valid in the heavy ($Q\bar q$) meson,
where $m_q\sim \omega$.

In the naive potential model, because the $m_{sp}^2/\omega^2$ ratio is
numerically close to 0.5, one can draw the conclusion about the approximate
validity of the meson dominance for the form factors of the
$B\to D^{(*)}l\nu$ transitions (see ref.\cite{12}). However, we find, that
the correct covariant description of the quark model leads to the
violation of the meson dominance for the transition form factors.

\section{Data Analysis and $|V_{bc}|$}

In the numerical analysis of the semileptonic decay widths,
in accordance to the optimal set of the parameters (set I)
\begin{eqnarray}
\bar \Lambda & = & 0.59\;\; GeV\;, \nonumber \\
m_q & = & 0.3\; GeV\;, \nonumber \\
m_b & = & 4.7\; GeV\;, \\
m_c & = & 1.4\; GeV\;, \nonumber \\
f_B & = & 80\; MeV\;, \nonumber \\
f_{D^{(*)}} & = & 130\; MeV\;, \nonumber
\end{eqnarray}
we have also considered set II
\begin{eqnarray}
m_q & = & 0.35\; GeV\;, \nonumber \\
m_b & = & 4.9\; GeV\;, \nonumber \\
m_c & = & 1.5\; GeV\;,  \\
f_B & = & 150\; MeV\;, \nonumber \\
f_{D^*} & = & 250\; MeV\;, \nonumber\\
f_D & = & 220\; MeV\;. \nonumber
\end{eqnarray}
The results of calculations for the total widths of the $B\to D^{(*)}l\nu$
decays are presented in Table \ref{t1} at $|V_{bc}|=0.038$.

\begin{table}[t]
\caption{The total widths of the $B\to D^{(*)}l\nu$ decays (in $10^{-6}$ eV),
calculated in the present model, in comparison with the experimental values
($<\tau_B>=1.5$ ps).}
\label{t1}
\begin{center}
\begin{tabular}{||c|c|c||}
\hline
Ref. & $B\to D l\nu$ & $B\to D^*l\nu$ \\
\hline
PDG\cite{20}    & $8\pm 3$   & $31\pm 10$ \\
set I  & $9.3$      & $30.6$\\
set II & $10.4$     & $30.5$  \\
\hline
\end{tabular}
\end{center}
\end{table}
In the limit of infinitely heavy quark, on the basis of expression (\ref{47})
for the Isgur-Wise function, the analysis of the differential
distribution in the decay $B\to D^*l\nu$ (see \cite{3,5,6,19}) leads
to the value
\begin{equation}
|V_{bc}| = 0.038\pm 0.003\;,
\end{equation}
that is in a reasonable agreement with the estimates in the other models
(see ref.\cite{19}).

\section{Discussion and Conclusion}

In the present paper we have considered the covariant potential model
of the heavy meson in the single-time prescription for the quark-meson
form factors with the modified choice of the relative quark momentum
inside the meson. In contrast to the naive potential model,
the offered consideration of the semileptonic $B\to D^{(*)}l\nu$ decays
is in an agreement with the experimental interval for the possible
values of the Isgur-Wise function slope, so that at the optimal choice
of the parameter for the oscillator-like wave function of the meson,
when the mass of the constituent light quark is adjusted with the width
of the wave packedge, one gets the value
$$
\rho^2 = 1.25\;,
$$
at which the analysis of the experimental data yields
$$
|V_{bc}| = 0.038\pm 0.003\;.
$$
Note, that the considered in \cite{11} Wigner rotation of the light quark
spin is beyond the applied approximation for the quark-meson form
factors and it is the attempt to take into account relativistic effects
of the light quark motion inside the meson.

We have also considered the conditions for the realization of the meson
dominance in the $q^2$-dependence of the semileptonic transition
form factors and shown, that the meson dominance can not take place
for the transitions of the heavy-light systems.

Thus, the offered quark model with the only parameter ($\bar \Lambda =
M-m_Q$) allows one reasonably to describe the QCD dynamics in the heavy meson
and to extract the $|V_{bc}|$ value from the data of the experimental
researches of the $B\to D^{(*)}l\nu$ decays. The reliability of this
description can be tested on the basis of the study of the data on
the other heavy meson decays\footnote{
This will be done elsewhere.}
and the improvement of the experimental
accuracy.


\begin{thebibliography}{**}
\bibitem{1}
M.A.Shifman, A.I.Vainshtein, V.I.Zakharov, Nucl.Phys. B147 (1979) 345,448;\\
E.V.Shuryak, Nucl.Phys. B198 (1982) 83;\\
T.M.Aliev and V.L.Eletskij, Yad.Fiz. 38 (1983) 1537 [Sov.J.Nucl.Phys. 38 (1983)
936];\\
L.J.Reinders, H.Rubinshtein and S.Yazaki, Phys.Rep. 127 (1985) 1;\\
C.A.Dominguez and N.Paver, Phys.Lett. B197 (1987) 423; (E) B199 (1987) 596;\\
S.Narison, Phys.Lett. B198 (1987) 104;\\
M.A.Shifman, Usp.Fiz.Nauk 151 (1987) 193 [Sov.Phys.Uspekhi 30 (1987) 91];\\
L.J.Reinders, Phys.Rev. D38 (1988) 947.
\bibitem{2}
A.A.Ovchinnikov, V.A.Slobodenyuk, Z.Phys. C44 (1989) 433;\\
V.N.Baier, A.G.Grozin, Z.Phys. C47 (1990) 669;\\
P.Ball, V.M.Braun, H.G.Dosch, Phys.Rev. D44 (1991) 3567;\\
A.F.Falk, H.Georgi, B.Grinstein, M.B.Wise, Nucl.Phys. B343 (1990)1;\\
A.V.Radyushkin, Phys.Lett. B271 (1991) 218;\\
M.Neubert, V.Rieckert, B.Stech, Q.P.Xu, Heidelberg Preprint HD-THEP-91-28,
1991;\\
E.Bagan, P.Ball, V.M.Braun and H.G.Dosch, Phys.Lett. B278 (1992) 457;\\
P.Ball, Phys.Lett. B281 (1992) 133.
\bibitem{3}
N.Isgur, D.Scora, B.Grinstein and M.B.Wise, Phys.Rev. D39 (1989) 799.
\bibitem{4}
M.Wirbel, B.Stech and M.Bauer, Z.Phys. C29 (1985) 637;\\
M.Bauer, B.Stech and M.Wirbel, Z.Phys. C34 (1987) 103;\\
T.Altomari and L.Wolfenstein, Phys.Rev. D37 (1988) 681;\\
J.K\" orner, G.Schuler, Z.Phys. C38 (1988) 511, C46 (1990) 93;\\
F.J.Gilman and R.L.Singleton, Phys.Rev D41 (1990) 142.
\bibitem{5}
H.Albrecht et al., ARGUS Collab., Phys.Lett. B275 (1992) 195,
Z.Phys. C57 (1993) 533.
\bibitem{6}
G.Crawford et al., CLEO Collab., CLEO-Conf. 93-30 (1993).
\bibitem{7}
N.Isgur and M.B.Wise, Phys.Lett. B232 (1989) 113, B237 (1990) 527.
\bibitem{8}
N.Isgur, Phys.Rev.D43 (1991) 810;\\
M.Neubert and V.Rieckert, Nucl.Phys. B382 (1992) 97;\\
E.de Rafael and J.Taron, Phys.Lett. B282 (1992) 215;\\
M.A.Ivanov, O.E.Khomutenko and T.Mizatuni, Phys.Rev. D46 (1992) 3817;\\
E.Jenkins and M.J.Savage, Phys.Lett. B281 (1992) 331;\\
A.Dubinin and A.Kaidalov, ITEP Preprint 71-92, Moscow, 1992.
\bibitem{9}
J.L.Rosner, Phys.Rev. D42 (1990) 3732;\\
T.Mannel, W.Roberts and Z.Ryzak, Phys.Lett. B254 (1991) 274;\\
M.Neubert, Phys.Lett. B264 (1991) 455.
\bibitem{10}
C.Bernard, Y.Shen and A.Soni, Phys.Lett. B317 (1993) 164.
\bibitem{11}
F.E.Close and A.Wambach, Nucl.Phys. B412 (1994) 169.
\bibitem{12}
V.V.Kiselev, Preprint IHEP 93-64, Protvino, 1994;
to appear in Int. J. Mod. Phys.A.
\bibitem{13}
E.Eichten et al., Phys.Rev. D21 (1980) 203.
\bibitem{14}
I.I.Bigi, M.A.Shifman, N.G.Uraltsev, A.I.Vainshtein, Preprint
CERN-TH. 7171/94 (1994).
\bibitem{15}
V.V.Kiselev, Preprints IHEP 94-75, 94-92, Protvino, 1994.
%\bibitem{16}
\bibitem{17}
Ch.-C.Chang and Y.-Q.Chen, Phys.Rev. D49 (1994) 3399.
\bibitem{18}
D.P.Stanly and D.Robson, Phys.Rev. D21 (1980) 3180;\\
C.Quigg and J.L.Rosner, Phys.Rev. D23 (1981) 2625;\\
A.Martin, Preprint CERN-TH.5349/88, Geneva, 1988;\\
D.B.Lichtenberg et al., Z.Phys. C47 (1990) 83;\\
W.Kwong and J.Rosner, Phys.Rev D44 (1991) 212;\\
S.S.Gershtein, V.V.Kiselev, A.K.Likhoded, S.R.Sla\-bo\-spit\-sky and
A.V.Tkab\-la\-dze,
Yad.Fiz. 48 (1988) 515 [ Sov.J. Nucl.Phys. 48(2) (1988) 327];\\
S.S.Gershtein, A.K.Likhoded and S.R.Slabospitsky,
Int.J.Mod.Phys. A6(13) (1991) 2309;\\
E.Bagan et al., Preprint CERN-TH.7141/94, 1994;\\
E.Eichten and C.Quigg, Phys.Rev. D49 (1994) 5845;\\
V.V.Kiselev, A.K.Likhoded and A.V.Tkabladze, Preprint IHEP 94-51, Protvino,
1994; to appear in Yad.Fiz.
\bibitem{19}
M.Danilov, Preprint ITEP 92-93, Moscow, 1993.
\bibitem{20}
L.Montanet et al., PDG, Phys.Rev. D50 (1994) 1173.
\end{thebibliography}
\end{document}